\documentclass[aps,prb,letterpaper,twocolumn,showpacs,superscriptaddress]{revtex4}

\usepackage{epsfig}
\usepackage{graphicx}
\usepackage{latexsym}

\usepackage{color}

\begin{document}


%
\title{Twins and their boundaries during homoepitaxy on Ir(111)}

%
%

\author{Sebastian Bleikamp}
\affiliation{{II}. Physikalisches Institut, Universit\"{a}t zu K\"{o}ln, 50937 K\"{o}ln, Germany}

\author{Johann Coraux}
\email[]{johann.coraux@grenoble.cnrs.fr}
\thanks{corresponding author}
\affiliation{Institut N\'{e}el, CNRS-UJF, 25 rue des Martyrs, BP166, F-38042 Grenoble Cedex 9, France and {II}. Physikalisches Institut, Universit\"{a}t zu K\"{o}ln, 50937 K\"{o}ln, Germany}

\author{Odile Robach}
\affiliation{Commissariat $\grave{a}$ l'Energie Atomique, Institut Nanosciences et Cryog$\acute{e}$nie, 17 Avenue des Martyrs, F-38054 Grenoble, Cedex 9, France}

\author{Gilles Renaud}
\affiliation{Commissariat $\grave{a}$ l'Energie Atomique, Institut Nanosciences et Cryog$\acute{e}$nie, 17 Avenue des Martyrs, F-38054 Grenoble, Cedex 9, France}

\author{Thomas Michely}
\affiliation{{II}. Physikalisches Institut, Universit\"{a}t zu K\"{o}ln, 50937 K\"{o}ln, Germany}

\date{\today}

\begin{abstract}

The growth and annealing behavior of strongly twinned homoepitaxial films on Ir(111) has been investigated by scanning tunneling microscopy, low energy electron diffraction and surface X-ray diffraction. \textit{In situ} surface X-ray diffraction during and after film growth turned out to be an efficient tool for the determination of twin fractions in multilayer films and to uncover the nature of side twin boundaries. The annealing of the twin structures is shown to take place in a two step process, reducing first the length of the boundaries between differently stacked areas and only then the twins themselves. A model for the structure of the side twin boundaries is proposed which is consistent with both the scanning tunneling microscopy and surface X-ray diffraction data.


\end{abstract}

%


\maketitle


\section{Introduction}

Stacking faults and twin boundaries are known to seriously affect the properties of thin films and devices. 
The boundaries with the initial crystallite along the twin plane is coherent, while the others, namely the side boundaries, are usually not. Incoherent boundaries disturb the thin film rather strongly and may give rise to pronounced property changes, \textit{e.g.} increase of the electrical resistivity. The magnetic properties of thin films are known to crucially rely on the stacking sequence  \cite{krause06,bautista07}. Coherent boundaries may act as electron scattering planes in metal nanowires \cite{wang04}, as traps for electrons in SiC  \cite{iwata01,lindefelt03} or to limit the performance of optoelectronic III-V nanowire-based devices \cite{johansson06}.

Due to the low energy of their formation, stacking faults are among the most frequent defects in thin films \cite{stowell75}. For the same reason, there is only a small driving force for their removal and twins are thus among the most stable defects. There is little knowledge on the kinetics of stacking fault healing at the atomic level, on temperature induced transformations of side boundaries, or on twin coarsening. 

Here we use homoepitaxy on Ir(111) to gain insight into the kinetics and atomistics of twin crystallite annealing in thin films. In a homoepitaxial system no stacking faults should be present due to energetics. Complications due to chemical inhomogeneities or epitaxial strain are absent. This article is a follow up of previous work on the formation and proliferation of stacking faults and twin crystallites in thin homoepitaxial Ir films on Ir(111) \cite{busse03,busse04,bleikamp06,bleikamp08,bleikamp09}. The present study combines the surface view of a partially twinned film obtained by low energy electron diffraction (LEED) and scanning tunneling microscopy (STM) with the surface+subsurface, quantitative view provided by surface X-ray diffraction (SXRD).

We find that the twinning in Ir thin films on Ir(111) gives rise to fractional step heights. Fractional step heights have already been observed several times on (111) surfaces of face-centered cubic (fcc) metals by scanning tunneling microscopy \cite{wolf91,lundgren00,christiansen02}. They result \textit{e.g.} from stacking fault tetrahedrons, Lomer-Cottrell locks or slip on $\left\{111\right\}$ planes inclined with respect to the $\left(111\right)$ surface. However, for all these cases the fcc stacking sequence is preserved on both sides of the fractional step, while in the present case of homoepitaxial films on Ir(111) fractional steps separate twinned and untwinned parts of the crystal.

Twin and stacking fault healing takes place in a two step process, in which first the amount of side boundaries is reduced through a crystallite coarsening process before eventually the amount of twin crystallites decreases. The geometry of the side boundaries is characterized by the SXRD measurements.

\section{Experimental}

The STM and LEED experiments have been performed in an UHV system with a base pressure below $5 \times 10^{-11}$\,mbar. 
The sample was cleaned by repeated cycles of sputtering with 1.5\,keV Xe$^+$ or Ar$^+$ ions at 1100\,K and annealing to 1600\,K. Iridium was deposited from a current heated Ir wire with a standard deposition rate of $1.3 \times 10^{-2}$\,ML/s, where 1\,ML (monolayer) is the surface atomic density of Ir(111).
Imaging was performed with a home built, magnetically stabilized STM \cite{michely00}. Unless otherwise specified, all images are differentiated and appear as illuminated from the left. The topographs were digitally post-processed with the WSxM software \cite{wsxm}. For the LEED I/V analysis, a rear view camera LEED system was used. 

The SXRD experiments have been performed at the SUV instrument\cite{BaudoingSavois99,BaudoingSavois99b} installed at the BM32 beamline at the European Synchrotron Radiation Facility (ESRF), in a UHV system, with a base pressure below $3\times 10^{-10}$\,mbar coupled to a Z-axis diffractometer. Sample preparation and deposition in this UHV system followed the same procedure as for the STM and LEED experiments. A monochromatic 18\,keV photon beam incident under an angle of $0.273^{\circ}$  with respect to the surface was used. The corresponding X-ray attenuation length for Ir is 51 {\AA}\cite{henke93} (23 ML). Note that this is an average value because the sample surface had a bowing of $0.2^{\circ}$ resulting in a spread of incident angle. The incident beam was doubly focussed to a size of 0.3 $\times$ 0.2\,mm$^2$ (full width at half maximum in horizontal and vertical directions respectively) at the sample location. Detector slits, located 570\,mm away from the sample, were set at 2\,mm parallel to the sample surface (which was vertical), and 2.2\,mm perpendicular to it (with 5\,mm guard slits at 200\,mm), resulting in an angular acceptance of $0.2^{\circ}$ in both directions.

%

For the general description of planes and directions, the standard cubic system is used ($a_{Ir}$ = 3.8392\,\AA). The surface plane is indicated by $\left(111\right)$ and the corresponding direction by $\left[111\right]$; side-facets of the same symmetry but not parallel to the surface are indicated by $\left\{111\right\}$ and the corresponding directions as $\left\langle  111\right\rangle$. 

\begin{figure*}
\begin{center}
\includegraphics[width=0.8 \linewidth]{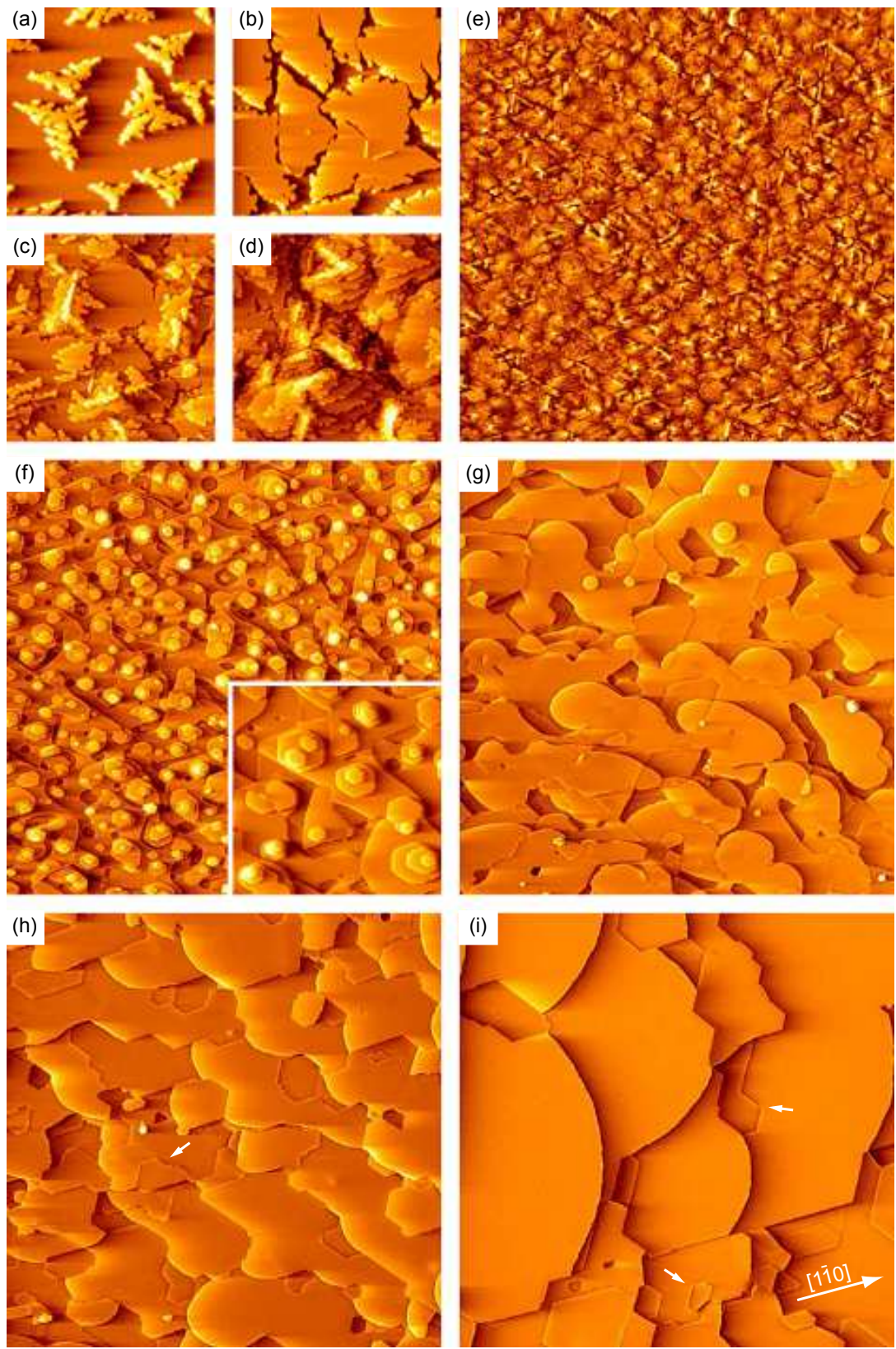}
\end{center}
\caption{
\label{uebersicht}
(Color online)  (a)-(d) STM topographs illustrating characteristic features at different stages of homoepitaxy on Ir(111) after deposition of (a) 0.2\,ML, (b) 0.9\,ML, (c) 10\,ML, (d) 90\,ML at 350\,K. Image size in (a)-(d) is 64\,nm\,$\times$\,64\,nm. (e) larger scale STM topograph of a 90\,ML film grown at 350\,K. (f)-(i) STM topographs of the film shown in (e) after succesive 180\,s annealing steps at (f) 850\,K, (g) 1025\,K, (h) 1200\,K, (i) 1375\,K. Image size in (e)-(i) is 480\,nm\,$\times$\,480\,nm. Inset size in (f) is 120\,nm\,$\times$\,120\,nm. Small white arrows in (h) and (i) indicate emerging screw dislocation lines (junctions between up and down steps).}
\end{figure*}

\section{STM experiments}

Fig.~\ref{uebersicht}(a)-(d) displays characteristic elements of a homoepitaxial growth sequence on Ir(111) already discussed in more detail previously \cite{bleikamp08,bleikamp06}. After deposition of 0.2\,ML as in Fig. \ref{uebersicht}(a) dendritic single atom high islands are visible. They display a triangular envelope due to the preferred formation of $\{111\}$-microfaceted step edges. The triangular shape of the island in the image center is mirrored (or rotated by $180^\circ$) in orientation. This mirrored orientation marks stacking fault islands with the atoms residing in the threefold coordinated hollow sites corresponding to an hcp rather than an fcc stacking sequence. Continued deposition gives rise to island coalescence and most stacking fault islands switch to regular stacking during this process \cite{busse04}. At the boundary between regular and remaining faulted areas narrow stripes of subatomic width remain, which offer fourfold coordinated adsorption sites. This gap and thus also the faulted area bounding on one of its sides become stabilized by adatoms forming a monatomic width decoration row oriented along $\left\langle 1\bar{1} 0\right\rangle $ [Fig.~\ref{uebersicht}(b)]. Continued growth takes place preferentially by heterogeneous nucleation at these decoration rows. It causes the formation of additional faults (fault proliferation \cite{bleikamp06}), stabilizes existing ones and leads to the embedding of twins. In Fig.~\ref{uebersicht}(c) the situation after 10\,ML deposited is shown, where overgrown decoration rows mark the separation of differently stacked surface areas. Eventually, further deposition gives rise to an irregular rough surface dominated by mounds. They result from heterogeneous nucleation at the boundaries between differently stacked areas of the films [compare Fig.~\ref{uebersicht}(d)].

The same film is shown in Fig.~\ref{uebersicht}(e) in a demagnified view. Each of the bright features marks a $\left\langle 1 \bar{1} 0\right\rangle$ oriented boundary between areas of different stacking\cite{footnote1}. In an annealing sequence with succesive annealing intervals of 180\,s this rough and heavily twinned film is subsequently heated to higher and higher temperatures. In Fig.~\ref{uebersicht}(f) after annealing to 850\,K the film has become much smoother and the typical structure size has increased.  Decoration rows or bright features oriented along $\left\langle 1\bar{1}0 \right\rangle$ are now absent. After annealing to 1025\,K [Fig.~\ref{uebersicht}(g)] the film is very flat and the characteristic structure size is further increased. Straight steps precisely oriented along the $\left\langle  1 \bar{1} 0 \right\rangle$ directions are visible. These straight steps have heights of only a fraction of a regular monatomic step. In addition, curved steps of monatomic height are visible. They are pinned where they touch fractional steps. The monatomic steps are invariably curved outward towards the downhill side. The outward curvature appears to imply an enhanced surface chemical potential, which for the case of Fig.~\ref{uebersicht}(g) may be traced back to decaying adatom islands also visible.
The last two annealing steps to 1200\,K [Fig. \ref{uebersicht}(h)] and 1375\,K [Fig.~\ref{uebersicht}(i)] increase the typical structure size further, \textit{i.e.} the step density decreases. The distinction between rounded monatomic steps and straight steps of fractional height is obvious. As indicated by arrows in Fig.~\ref{uebersicht}(h) and (i) certain areas bounded by fractional steps are misoriented with respect to the global $\left[111\right]$ orientation. This is apparent through twisted steps being partly up and partly down steps which indicates the presence of dislocations in the film.
The fact that even in Fig.~\ref{uebersicht}(i), after the decay of all adatom islands, pinned monatomic steps are invariably curved outwards is possibly caused by less strongly bound atoms in the defect structures associated with the fractional steps. The defect structures define an enhanced chemical potential visible through the step curvature.

\begin{figure*}
\begin{center}
\includegraphics[width=1.0 \linewidth]{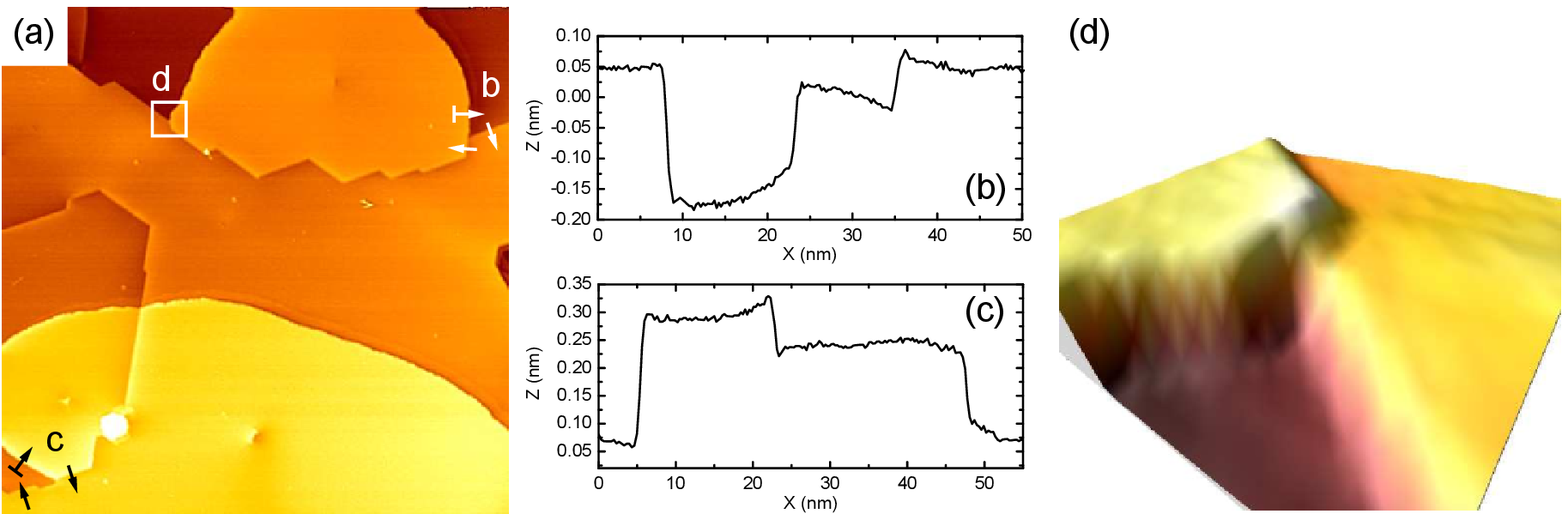}
\end{center}
\caption{
\label{partial}
(Color online)  (a) Greyscale STM topograph after deposition of 90\,ML film at 350 K and additional annealing to 1025\,K for 180\,s. Image size is 128\,nm\,$\times$\,128\,nm. (b), (c) Line scans along paths indicated in (a) illustrating full and fractional step heights as well as strain fields associated with fractional steps. (d) Three dimensional visualization of the junction marked in (a).}
\end{figure*}

Fig.~\ref{partial}(a) shows a height image (not differentiated) of an area with straight (fractional) and curved (monatomic) steps in detail.
Example line scans around the junction points marked with the arrows in (a) are shown in Fig.~\ref{partial}(b) and (c). The situation at a typical triple junction is depicted in a three dimensional view of Fig.~\ref{partial}(d).
Neglecting for the moment heavily distorted areas, only three different kind of steps are found: (i) steps with a height of 2.22 \AA \, $\pm$ 0.17 \AA, i.e steps with a height corresponding to the (111) layer distance of 2.22\,\AA; (ii) steps with a height of 0.74 \AA\, $\pm$ 0.09 \AA\, \textit{i.e.} steps which display within the limits of error a fractional height of $1/3$ of a monatomic step; (iii) steps with a height of 1.49 \AA\, $\pm$ 0.24 \AA\, \textit{i.e.} steps which display within the limits of error a fractional height of $2/3$ of a monatomic step [compare Fig.~\ref{partial}(b) and (c)]. The measured step heights do not depend on tunneling voltage and are thus a topographic effect.
Evidently, fractional steps cannot be present on the (111) surface of a perfect Ir crystal and their occurrence proves the presence of extended defect structures in the Ir film even after annealing to 1375\,K [compare Fig.~\ref{uebersicht}(i)]. 
In the vicinity of the fractional steps, typically an upward (downward) bending of the upper (lower) terrace perpendicular to the step direction takes places. It extends over a distance of about 9\,nm  with a significant scatter and indicates the presence of strain next to the fractional steps.

\begin{figure}
\begin{center}
\includegraphics[width=0.999 \linewidth]{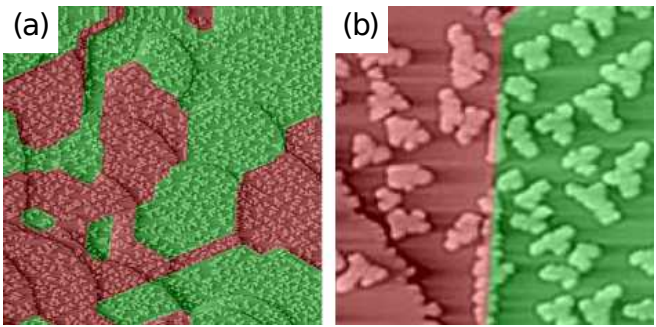}
\end{center}
\caption{
\label{markierte}
(Color online) STM topographs after deposition of 90\,ML at 350\,K, subsequent annealing to 1200\,K for 180\,s and final 0.2\,ML Ir deposition at room temperature. The predominant orientation of the triangular envelope of the small dendritic islands allows one to identify the stacking sequence of the underlying film. fcc areas are shaded green, faulted areas are shaded red. Image sizes are (a) 255\,nm\,$\times$\,255\,nm and (b) 44\,nm\,$\times$\,44\,nm.}
\end{figure}

A post decoration technique was employed to assess the stacking of the surface areas of the Ir films after annealing \cite{meinel88}. 0.1-0.2\,ML Ir were deposited on the annealed film surface at room temperature. Dendritic Ir islands with the characteristic triangular envelope grow, which display the same orientation in an area of uniform stacking with a preference of about 90 \%. Areas of mirrored stacking display consequently mirrored preferences of the triangular island envelopes. Application of the post decoration method to the surface after annealing at 1200\,K is exemplified in Fig.~\ref{markierte}. The areas have been marked depending on the stacking sequence identified. We find that even after annealing to 1200\,K more than 50 \% of the surface area of the 90\,ML films is twinned with respect to the bulk crystal. This result is backed up by quantitative LEED I/V analysis \cite{bleikamp08} and further supported in this article by quantitative analysis of SXRD measurements which are sensitive to the stacking of the film grains over their full thickness. 

As illustrated in Fig.~\ref{markierte}(b), fractional steps separate areas of different stacking sequence.
In fact, for an annealed film adjacent areas of different stacking are always separated by a fractional step. 
Thus fractional steps mark the lateral border of the differently stacked grains normal to the twin plane, \textit{i.e.} they indicate the positions of side twin boundaries. The orientation of the separating steps is along the $\left\langle 1\bar{1}0 \right\rangle$ which is also the intersection between $\{111\}$ planes and the (111) surface. As these fractional steps separate differently stacked grains, the planar defects associated with them must be different from the cases of fractional step heights analyzed previously by STM \cite{wolf91,lundgren00,christiansen02}. In order to understand the exact nature of these boundaries, sub-surface information is required, which SXRD provides us with.

\section{SXRD experiments}

\begin{figure}
\begin{center}
\includegraphics[width=0.9 \linewidth]{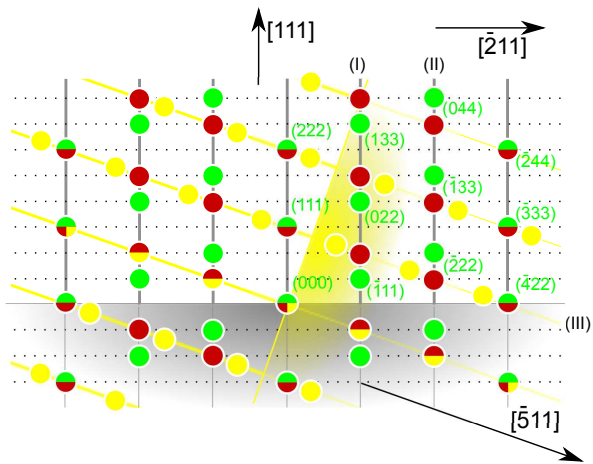}
\end{center}
\caption{
\label{ctrmap}
(Color online) Reciprocal space cut in the plane defined by the $[111]$ and $[\bar{2}11]$ directions of an fcc crystal. The medium grey (green) disks depict the reciprocal lattice of a perfect fcc crystal; the dark grey (red) disks depict the reciprocal lattice of a twin crystal, \textit{i.e.} of an fcc crystal rotated by 180$^\circ$ around a $[111]$ axis; the light grey (yellow) disks depict the reciprocal lattice of this twinned crystal, but additionally rotated by 180$^\circ$ around the $[\bar{5}11]$ direction of the initial crystal (twin-of-twin) which matches the $[\bar{1}11]$ direction of the twin crystal. For multi-twin peaks, half- or quarter-disks are employed. Shaded areas represent the truncated crystals: the grey crystal is terminated by a $(111)$ surface while the yellow one corresponds to the $(\bar{1}11)$ termination of the twin, \textit{i.e.} a $(\bar{5}11)$ surface in the frame of the initial crystal. These truncated crystals produce CTRs which are displayed as solid lines (grey or yellow depending on the crystal).}
\end{figure}

Figure~\ref{ctrmap} displays the reciprocal lattice of a perfect (bulk) fcc crystal (green disks), cut in the plane perpendicular to $(111)$ and containing the $[\bar{2}11]$ direction. For a truncated crystal, the loss of symmetry in direct space creates a reciprocal lattice of crystal truncation rods [CTRs, solid lines in Fig.~\ref{ctrmap}] perpendicular to the surface and connecting the reciprocal lattice points corresponding to the bulk crystal \cite{Robinson1986}. Accordingly the main intensity maxima coincide with the fcc bulk reflections. We now consider Ir homoepitaxial growth. As we have shown previously \cite{busse03,busse04,bleikamp06,bleikamp08,bleikamp09}, the overgrown Ir may either prolong the fcc stacking, or grow with a stacking fault initiating the growth of a twinned crystal portion, \textit{i.e.} a crystal portion rotated by 180$^\circ$ around a $[111]$ axis. The first stacking contributes to the CTRs intensities in a similar way as the substrate does, \textit{i.e.} with intensity maxima for the fcc bulk reflections. The second stacking gives a reciprocal lattice rotated by 180$^\circ$ around $[111]$. As an illustration, for the CTR passing through $1/3 (\bar{4}22)$ [labeled (I) in Fig.~\ref{ctrmap}], intensity maxima at $(\bar{1}11)$, $(022)$, $(133)$, etc, corresponding to the fcc stacking (green disks), are supplemented by maxima shifted with $1/3 [111]$, corresponding to the twinned fcc stacking (red disks). Monitoring the relative fcc and twinned intensities along the CTRs during growth qualitatively informs on the relative fcc/twinned fraction,\cite{camarero00} the signal of the film parts with fcc stacking being initially difficult to separate from the much higher superimposed signal due to the substrate. As extensively shown in the literature for the case of a single or a few atomic layers, more careful analysis, noticeably accounting for interferences between the twinned and untwinned regions, allows one to quantitatively determine the stacking composition in these layers. Below we show that such quantitative analysis can also be performed for much thicker films ($\simeq$70 atomic layers).

\begin{figure*}
\begin{center}
\includegraphics[width=.6 \linewidth]{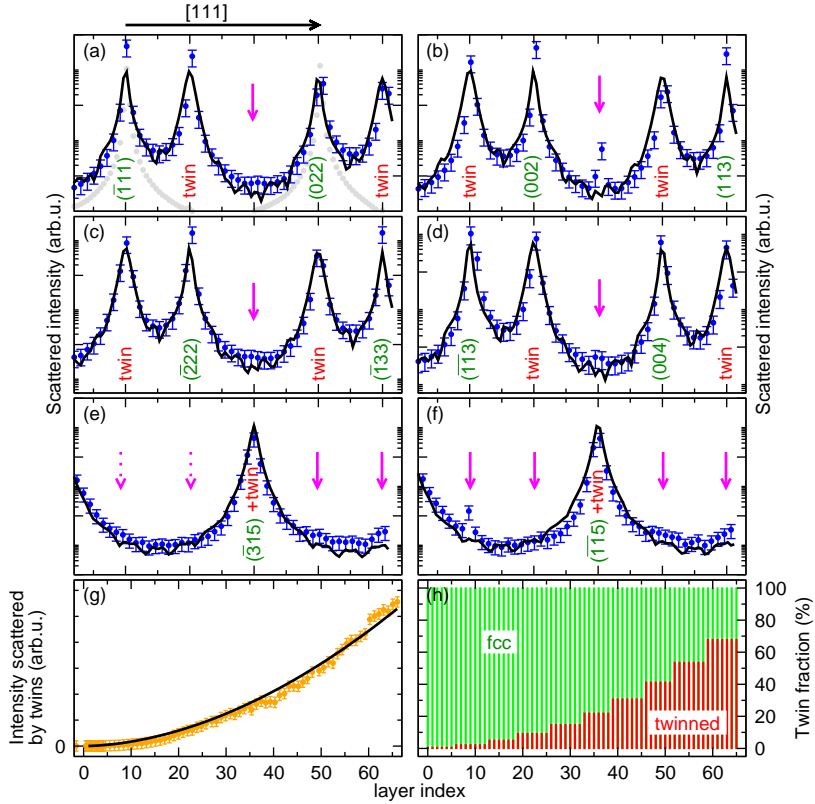}
\end{center}
\caption{
\label{ctrs}
(Color online) (a-f) Integrated intensity scattered along six CTRs perpendicular to the $(111)$ surface, \textit{i.e.} along the $[111]$ direction, after deposition of 68\,ML Ir on Ir(111) at 350\,K. Dots represent the data, while solid lines correspond to the fit of the whole set of CTRs. (a) and (c) correspond to the CTRs marked (I) and (II) in Fig.~\ref{ctrmap}. The Bragg reflections for the fcc untwinned crystal are indicated in green. Positions for the twinned crystal are indicated by "twin". The vertical arrows point to positions where faint extra contributions are observed. The data set displayed in grey in (a) is the integrated scattered intensity for the substrate prior to the growth of the Ir thin film. (g) Integrated intensity at the left position marked "twin" in (b), as a function of the deposit amount. The dots represent the data, the solid line was computed through the parameters from the fit of the CTRs. (h) Fraction of twin/fcc regions in the Ir overgrown film, as a function of the layer index, as deduced from the fit of the CTRs.}
\end{figure*}

Figure~\ref{ctrs}(a-f) show the X-ray scattered intensity along a set of six CTRs, all of them along the $[111]$ direction [Figs.~\ref{ctrs}(a,c) are the CTRs marked (I) and (II) in Fig.~\ref{ctrmap}, the others could be sketched in other appropriate cuts in reciprocal space], after deposition of 68\,ML of Ir at 350\,K. The occurrence and location of strong maxima different from the bulk fcc reflections along  four of the CTRs [Figs.~\ref{ctrs}(a-d)], together with the absence of noticeable extra maxima along the two others [Figs.~\ref{ctrs}(e,f)], are an unambiguous signature for a large twin fraction in the Ir thin film. In between the Bragg peaks, the intensity variations correspond to the roughness only for Fig.~\ref{ctrs}(e) and to a mixture of the roughness contribution and of the twin signal for Figs.~\ref{ctrs}(a,c). The Bragg peaks for the substrate [on the grey curve in Figs.~\ref{ctrs}(a)] are narrower than those from the fcc regions in the thin films, confirming that the X-ray beam probes prominently the top region of the sample where the effective height of the fcc domains is reduced -- as compared to the case for the bare substrate -- due to the presence of twin domains. The appearance of the twin signal was monitored \textit{in situ} during the growth by measuring the integrated intensity around the left position marked "twin" in Fig.~\ref{ctrs}(b). The result is shown in Fig.~\ref{ctrs}(g), where a smooth increase is observed starting from zero. Note that in contrast to the STM analysis, this evolution is not only taking into account the fraction of twins at the topmost layer, but also the twin in the layers below as a result of the penetration depth of X-rays. A quantitative analysis, presented in the next paragraph, will allow us to determine the Ir film stacking composition profile and roughness. In addition to the typical twin signals, fainter extra features are also observed, to more or less extent, along the CTRs. These features are marked by vertical arrows in Figs.~\ref{ctrs}(a-d). Some of them are very faint, two of them are even barely seen. Their observation, though very sensitive to the alignment of the goniometer axis, is confirmed in a second series of experiments. Their origin is discussed latter.

The quantitative analysis of the Ir film composition was performed by using a simple model which is detailed in appendix A, allowing us to simulate the scattered intensity along the whole set of CTRs with a single set of a few parameters that are refined by a least squares minimization algorithm. Through this procedure we obtained the thin film stacking composition in terms of twin/fcc fractions as a function of the height in the film as displayed in a graphical way in Fig.~\ref{ctrs}(h). The model finds a twinned fraction of ~70 \% at the surface, which agrees, within the limits of uncertainties, with the STM data presented latter. We also obtained an rms roughness of 4.3\,\AA~(assumed identical for twinned and fcc regions), for which a comparison to the data obtained through the STM analysis (2.5\,\AA) is hazardous (see Appendix A).

The large amount of time needed for the measurement of a full set of CTRs prevented us from doing such a measurement during the growth of the thin film. As an alternative we measured the integrated intensity over a restricted range in reciprocal space, below a twin peak [Fig.~\ref{ctrs}(g)] and an fcc peak as a function of the film thickness, which \textit{a priori} only provides us with a qualitative characterization of the evolution of the amount of twins. This evolution may however be compared to the scattered intensity calculated with a model Ir thin film of increasing thickness which stacking composition follows the quantitative picture developed as described in the previous paragraph. The calculation nicely reproduces the data as shown in Fig.~\ref{ctrs}(g), which further confirms the validity of the quantitative picture of the thin film. The evolution displayed in Fig.~\ref{ctrs}(g) may then be interpreted by inspection of our model  [Fig.~\ref{ctrs}(h)]. After deposition of about 10\,ML the twin fraction becomes noticeable and increases faster and faster. We know from STM experiments \cite{notonroughness}, that after 10\,ML  the growth mode changes from layer-by-layer growth to rough growth dominated by heterogeneous nucleation at decoration rows [compare Fig.~\ref{uebersicht}(c)] with rapid proliferation of stacking faults.

\begin{figure*}
\begin{center}
\includegraphics[width=0.6 \linewidth]{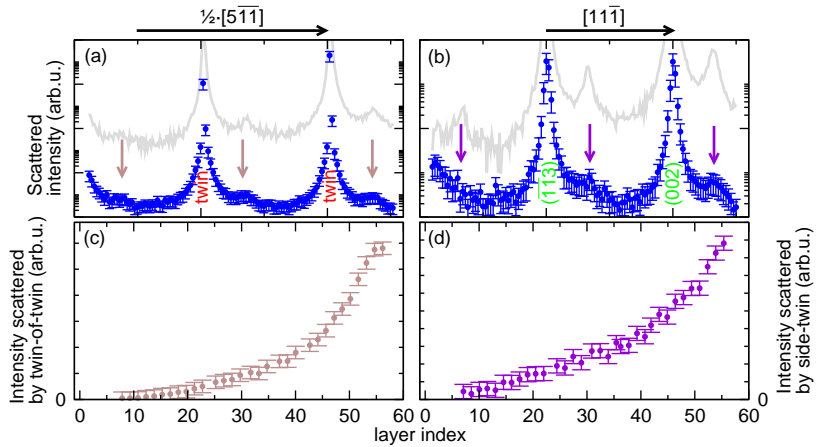}
\end{center}
\caption{
\label{ctrs_alt}
(Color online) (a,b) Intensity scattered along lines in the $[5\bar{1}\bar{1}]$ (a) and $[11\bar{1}]$ (b) directions, after deposition of 68\,ML Ir on Ir(111) at 350\,K. (a) corresponds to the CTRs marked (III) in Fig.~\ref{ctrmap}. The Bragg reflections for the fcc untwinned crystal are indicated in light grey (green), those for the twinned crystal are marked by "twin". The vertical arrows point to positions of the twins-of-twins (a) and of the side-twins (b), which are better visible for the scans (grey lines, shifted upwards for clarity) corresponding the scattered intensity after annealing at 750\,K. The intensity measured off the peaks is a mixture between the side facets [$(5\bar{1}\bar{1})$ and $(11\bar{1})$] signals and that from the corresponding stacking faults. (c,d) Intensity at the central position marked by a vertical arrow in (a),(b) as a function of the Ir deposit (see text).}
\end{figure*}

We then explored the possible stacking sequences on the side facets of fcc and twin Ir islands, \textit{i.e.} on their $\{111\}$ facets other than $(111)$  (for twin Ir islands these are $\{115\}$ planes if expressed in the fcc frame), and the possible appearence of new Ir orientations corresponding to twinning with these inclined facets as twin planes. This leads to six new possible twin-daughter orientations of Ir, three from the fcc parent and three from the $(111)$-twin parent. Accordingly we explored lines in reciprocal space being perpendicular to the $\{111\}$ and $\{115\}$ planes respectively. Figure~\ref{ctrs_alt}(a) shows the integrated scattered intensity along the line marked (III) in Fig.~\ref{ctrmap}, \textit{i.e.} along the $\left[5\bar{1}\bar{1}\right]$ direction which is normal to a  $\{111\}$ plane of a twin crystallite. In addition to the expected twin peaks, new ones develop as highlighted by the vertical arrows and better visible on the grey curve which was obtained after annealing. The intensity of such peaks was monitored during growth [Fig.~\ref{ctrs_alt}(c)]. These new peaks are generated by twin daugthers from the $(111)$-twin parent (twin-of-twin), as can be seen in Fig.~\ref{ctrmap}. A similar observation is made along a line lying perpendicular to a $(11\bar{1})$ plane [Figs.~\ref{ctrs_alt}(b,d)]: the fcc reflections are supplemented by extra features developing during growth which are generated by one of the twin-daughter from the fcc parent (side-twin), more precisely the one resulting from a 180$^\circ$ rotation around the $[11\bar{1}]$ direction.

\section{Temperature dependence}

\begin{figure}
\begin{center}
\includegraphics[width=0.999 \linewidth]{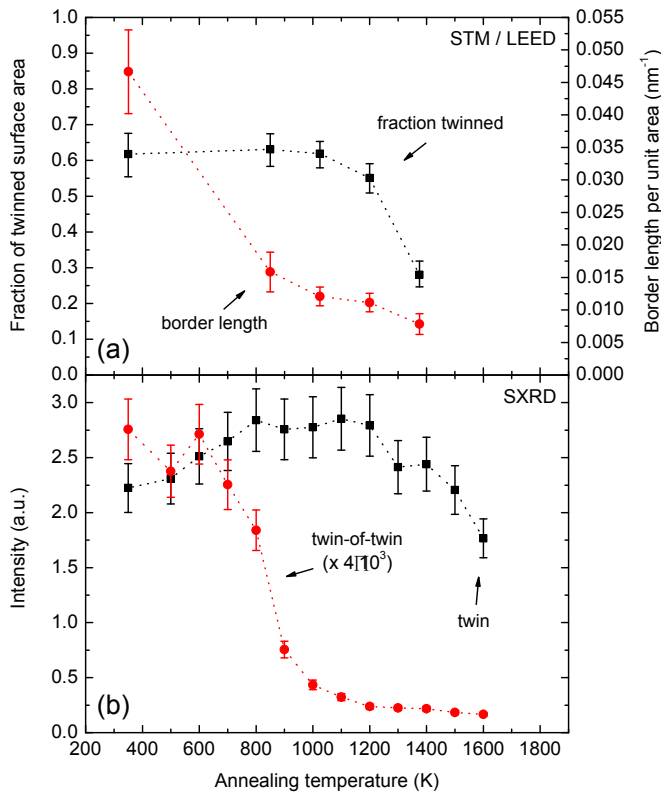}
\end{center}
\caption{
\label{ausheilkurven}
(Color online) 
(a) Fraction of twinned surface area measured by LEED (black squares) and border length between differently stacked surface areas measured by STM (red circles) for a 90\,ML Ir film grown on Ir(111) at 350\,K, as a function of the annealing temperatures.
(b) Intensities from the twin [black squares, measured at the left position marked "twin" in Fig.~\ref{ctrs}(b)] and the twin of twin [red circles, measured at the central location marked with a vertical arrow on Fig. \ref{ctrs_alt}] as a function of the annealing temperature.}
\end{figure}

Figure~\ref{ausheilkurven} compares the annealing temperature dependence of twin fraction estimated by the different techniques. The black squares in Fig.~\ref{ausheilkurven}(a) display the fraction of twinned surface area as a function of annealing temperature measured by LEED. It is around 0.6 up to 1025\,K, then gradually drops but does not vanish even for the highest annealing temperature of 1400\,K. The integrated X-ray intensity around the left position marked "twin" in Fig.~\ref{ctrs}(b), shown in Fig.~\ref{ausheilkurven}(b) as black squares, is also indicative of twinning with respect to the $\left( 111 \right)$ surface plane. It does not only measure the surface area but also the average twin fraction in the film volume underneath the surface due to the integration depth of SXRD. The amount of twinning apparently does not change up to an annealing temperatures of 1200\,K and only then gradually decreases. Roughly, both data sets agree that (i) twins start to decay in the temperature range of 1000\,K to 1200\,K and that (ii) even the highest annealing temperature used is not sufficient to remove them entirely. We attribute the shift in the onset of twin annealing towards higher temperature in the X-ray data to the higher pressure during Ir thin film deposition. In LEED annealing sequences we noticed that this onset depends sensitively on the pressure during deposition. Most likely it is related to a higher amount of carbon containing residual gas, which partly decomposes on the surface during growth. This results in incorporation of carbidic species into the film, which may hinder twin annealing at higher temperatures.

The full dots (red) in Fig.~\ref{ausheilkurven}(a) show the length of the borders separating surface areas of different stacking sequence (side twin boundaries) as obtained by STM. At low temperature the border length is derived form decoration rows and ridges in mounds elongated along $\left\langle 110\right\rangle$ and at high temperature from the length of the fractional steps. The border length decreases rapidly upon annealing at temperatures which are well below the onset temperature for a decrease of  the fraction of twinned surface area [full squares (black) Fig.~\ref{ausheilkurven}(a)]. Apparently both twinned and untwinned areas become more compact (reducing side twin boundaries), and probably even grow in size through coalescence processes or preferential disappearance of small domains. Note that these processes take place without affecting the relative fractions of twinned and untwinned surface area. This is plausible, as the driving force to remove side twin boundaries (separating laterally areas of different stacking) is much larger than the driving force to remove the coherent twin boundaries: grain boundary energies are much higher than stacking fault energies. The full red dots in Fig.~\ref{ausheilkurven}(b) measure the X-ray intensity at the central position marked by a vertical arrow in Fig.~\ref{ctrs_alt}(a), \textit{i.e.} for twins on a $\left\{111\right\}$ side-facet of a twinned crystallite, during annealing. The measured intensity displays a similar temperature dependence as the border length measured by STM. It decreases already around 700\,K, at much lower temperatures than the twin intensity itself. We therefore conclude that the \emph{twin-of-twin} intensity is linked to the side twin boundaries separating patches of different stacking.

\section{Discussion}

Below we develop a picture of the twinned film to account for the following facts:
(i) the boundaries between areas of different stacking sequence are parallel to the $\left\langle 1 \bar{1} 0 \right\rangle$ directions;
(ii) they are separated by fractional steps of 1/3 or 2/3 height;
(iii) twins with respect to the $\left\{ 111\right\}$ side facets of twinned and untwinned crystallites are linked to the boundaries between differently stacked regions;
(iv) weak unexpected peaks along the CTRs parallel to $[111]$ [vertical arrows in Figs.~\ref{ctrs}(a-f)] are generated during the growth.

\begin{figure}
\begin{center}
\includegraphics[width=0.8 \linewidth]{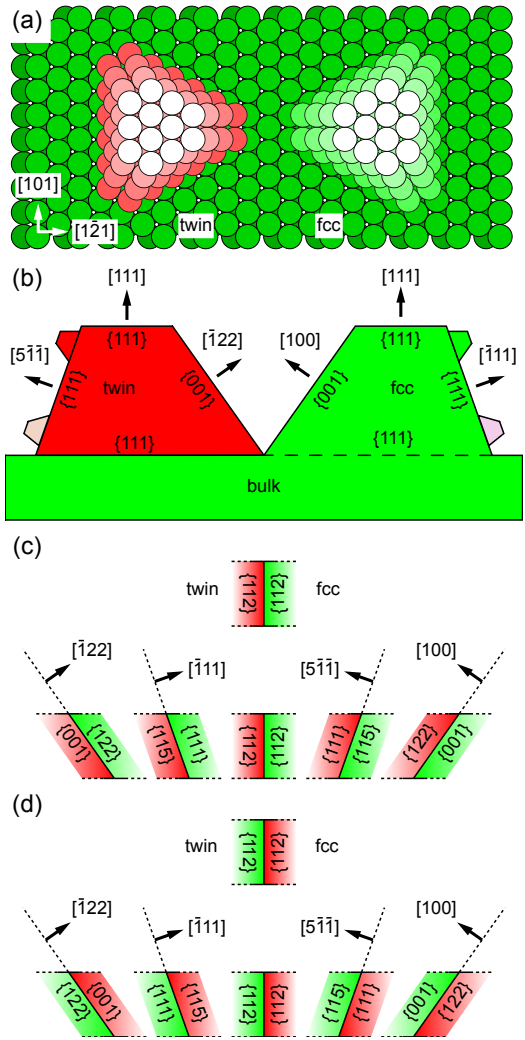}
\end{center}
\caption{
\label{grenzen}
(Color online) 
(a) Ball model of an fcc and a twin crystallite bounded by $\left(111\right)$ top facets and $\left\{111\right\}$ as well as $\left\{100\right\}$ side-facets.
(b) Cross section depicting the facets and the corresponding orientations with respect to the regular crystal matrix. On the left (right) facet of the twin (fcc) crystallite are shown two small crystallites, one with the same structure as the initial twin (fcc) crystallite which is displayed with the same colour, and one which structure is twinned with respect to the initial twin (fcc) crystallite. The two latters are the twin-of-twin and side-twin and appear in light grey (light brown and violet).
(c),(d) Models of possible side twin boundaries involving low index planes.
}
\end{figure}

A key to such a picture is the nature of the side twin boundaries. Fig.~\ref{grenzen}(a) displays an untwinned and a twinned crystallite bounded by $\left(111\right)$ top facets and $\left\{111\right\}$ as well as $\left\{100\right\}$ side-facets. A cross section through the crystallites indicating the facets and their directions is shown in Fig.~\ref{grenzen}(b). Note that the directions are given in units of the substrate lattice vectors. Therefore, the $\left[\bar{1}12\right]$ direction is not only normal to a $\left\{\bar{1}12\right\}$ side-facet of a fcc crystallite, but also to a $\left\{100\right\}$ side-facet of a twinned crystallite. Figs.~\ref{grenzen}(c) and (d) sketch possible crystal plane combinations with low index planes forming side twin boundaries. To embed a twin crystallite into an untwinned matrix (or the opposite) one of the boundaries sketched in Fig.~\ref{grenzen}(c) and one sketched in Fig.~\ref{grenzen}(d) are necessary. The principal plane combinations are of the types $\{112\}/\{112\}$, $\{001\}/\{122\}$ and $\{111\}/\{115\}$.

\begin{figure}
\begin{center}
\includegraphics[width=1.0 \linewidth]{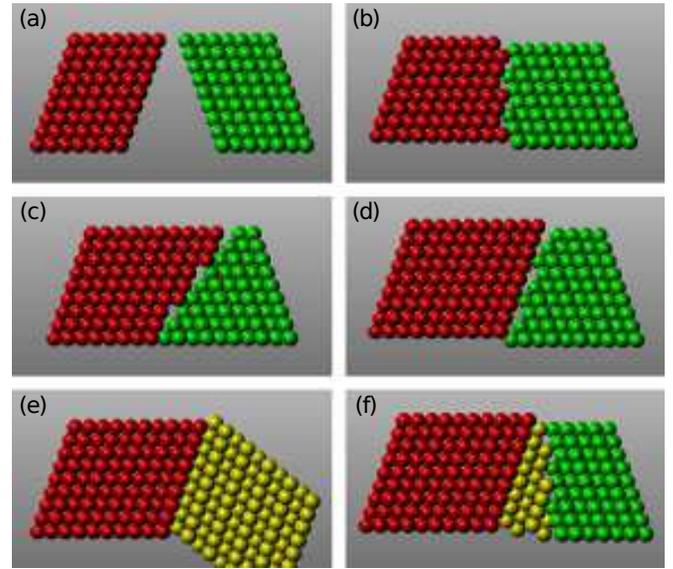}
\end{center}
\caption{
\label{structure}
(Color online) 
Ball models for high symmetry side twin boundaries.
(a) Orientation of the crystallites: (red/left) twinned, (green/right) fcc.
(b) Model of a $\{112\}/\{112\}$ twin boundary with partial dislocation 1/2 $[111]$.
(c) Model of a $\{001\}/\{122\}$ boundary with a partial dislocation of 1/2 $[ 1 \bar 1 0 ]$.
(d) Model of a $\{111\}/\{115\}$ boundary with a partial dislocation of 1/3 $[255]$.
(e) Model of a crystal (yellow) twinned to a twin crystal (red) at a $\left\{111\right\}$ facet.
(f) Final model for the side twin boundary (see text).}
\end{figure}

The $\{112\}/\{112\}$  twin boundary is normal to the sample surface. As known from TEM investigations of other fcc metals, \textit{e.g.} Au,\cite{marquis07} the $\{112\}/\{112\}$ boundary leads to a partial dislocation of 1/2 $\left[111\right]$ as shown in Fig.~\ref{structure}(b), \textit{i.e.} the atomic positions in the twin crystal are deduced from the ones of the fcc crystal by a rotation of 180$^\circ$ around $\left[111\right]$, followed by a translation of 1/2 $\left[111\right]$. The addition of this translation allows a closer contact between the two crystals along the $\{112\}/\{112\}$ boundary. This would result in a fractional step height of 1/2 monolayer in contradiction to our STM observation of 1/3 and 2/3 fractional steps.

In a $\{001\}/\{122\}$ boundary, every third atom of the $\{122\}$ facet is in contact with the $\{001\}$ facet and rests in the fourfold coordinated sites formed by the atoms of the $\{001\}$ facet [Fig. \ref{structure}(c)]. This is obtained by adding a translation along 1/2 $\left\langle 1 \bar 1 0 \right\rangle$, \textit{i.e.} in the surface plane, to the 180$^\circ$ rotation. The surfaces of the regular and twinned parts stay at the same height and thus this type of boundary is inconsistent with the observed partial step heights.

The $\{111\}/\{115\}$ boundary [Fig. \ref{structure}(d)] has one third of the atoms of the $\left\{115\right\}$ plane in contact with the $\{111\}$ facet. A translation of 1/3 $\left\langle 211 \right\rangle$, \textit{i.e.} in the $\{111\}$ facet, makes the atoms in contact sit in threefold coordinated fault sites of the $\left\{111\right\}$ facet and increases the density of this boundary considerably. The same holds for the twin side-facets, and corresponds to a translation of 1/3 $\left\langle 255 \right\rangle$, \textit{i.e.} in the $\{115\}$ facet. These configurations involve characteristic steps of 1/3 or 2/3 of a monolayer step height. 

\begin{figure*}
\begin{center}
\includegraphics[width=0.6 \linewidth]{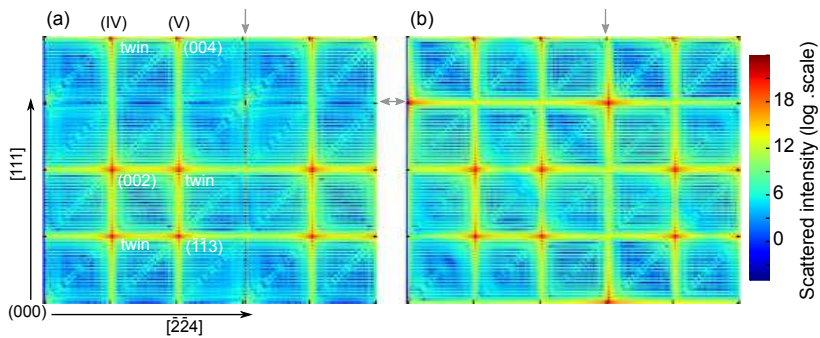}
\end{center}
\caption
{(Color online)  Intensity map for a $\{111\}/\{115\}$ boundary between an untwinned and a twinned crystal of the same size, (a) without and (b) with a shift on the $(11 \bar 1)$ side-facet resulting in a height difference of 1/3 monolayers (see text). The fringes are due to the finite size of the crystallites considered for the calculation.}
\label{imap115}
\end{figure*}

We simulated the X-ray scattered intensity in three dimensions, for a system consisting of two crystallites, one fcc, and the other $(111)$-twin, bounded by a $\{111\}/\{115\}$ twin boundary [a $(11\bar{1})$ side-facet on the fcc side]. Details of the calculation are given in appendix B. Two situations were considered: (i) the atoms of the first plane for each crystallite are located at positions corresponding to hollow sites for the $(111)$ surface, in hcp sites for the twinned part and in fcc sites for the fcc part, and the positions of all atoms above are derived accordingly; (ii) from (i), the twinned crystallite is shifted by 1/3 $[255]$ as in Fig.~\ref{structure}(d). The calculation along the reciprocal space plane defined by $[111]$ and $[\bar{1}\bar{1}2]$ is shown in Figs.~\ref{imap115}(a) and (b) for (i) and (ii) respectively. The most obvious difference between the two maps is the appearance of two lines, one vertical, the other horizontal (grey arrows in Fig.~\ref{imap115}) along which constructive interferences yield increased intensity. The horizontal feature intersects all $[111]$-CTRs, including also those scanned in Figs.~\ref{ctrs}(b,d) [marked (IV) and (V) respectively in Fig.~\ref{imap115}]. It is worth noting here that the intensity is not much increased right at the intersection between (IV), (V), and the horizontal feature. However, if, due to some small misalignment, the CTRs are scanned a little off their exact position, the extra feature will become better pronounced. Due to limited goniometer alignment, this is observed in Figs.~\ref{ctrs}(b,d) to a certain extent. Similar features were reproduced as well for all other cuts of the reciprocal space relevant to the CTRs scanned in Figs.~\ref{ctrs}(a-d). The extra features observed on the CTRs are thus presumably generated by the 1/3 $[255]$ shifts and therefore the 1/3 and 2/3 steps.

Fig. \ref{structure}(d) provides also some clue about a possible origin of the strain fields accompanying the steps of fractional height apparent \textit{e.g.} in Fig.~\ref{partial}(a). In a certain distance from the $\{111\}/\{115\}$ boundary both crystallites must be attached to the same plane provided by the substrate - otherwise an energetically costly extended void would result. Attachment of both crystallites to the same substrate implies lattice deformations in the vicinity of the $\{111\}/\{115\}$ boundary.    

We now address the side-twin and twin-of-twin peaks appearing along the lines parallel to $[5\bar{1}\bar{1}]$ and $[11\bar{1}]$ [Figs.~\ref{ctrs_alt}(a,b)]. The crystallites with densest facets corresponding to these two structures are shown in Fig.~\ref{grenzen}(b). The structure corresponding to the twin-of-twin is displayed in Fig.~\ref{structure}(e). Although this accounts for the X-ray measurements, the $(111)$ facet of such a doubly twinned domain [shown in yellow in Fig.~\ref{structure}(e)] would have a large inclination of $39^\circ$ [see the facet of the twin-of-twin crystallite in Fig.~\ref{grenzen}(b)] degrees with respect to the $\left(111\right)$ facet of twinned and fcc crystallites and thus would intersect the thin film surface and would be seen in the STM topographs. We find no evidence for such inclined facets at the surface.

A reason for this could be that the $\{111\}/\{115\}$ boundary dissociates into $\{111\}/\{111\}$ [Fig.~\ref{structure}(e)] and $\{552\}/\{112\}$ [Fig.~\ref{structure}(b)] boundaries, as sketched in Fig.~\ref{structure}(f). This is well supported by the literature, where the instability and dissociation of the $\{111\}/\{115\}$ boundary is reported \cite{merkle92}. Note that such a dissociation still assumes that the twin and fcc domains are translated with respect to each other with a 1/3 $[255]$ shift which is responsible for the extra feature discussed earlier. Energetically, the dissociation proposed is consistent with the available literature data for other fcc systems. For instance Wolf \textit{et al.}\cite{wolf92} calculated for Cu the energy of a $\{111\}/\{115\}$ boundary to be 0.61 Jm$^{-2}$, \textit{i.e.} larger than the sum of the energies for a $\{111\}/\{111\}$ boundary (0.01 Jm$^{-2}$) and for a $\{112\}/\{552\}$ boundary (0.26 Jm$^{-2}$).

The bottom facet of the light grey (yellow) crystallite visible in Fig.~\ref{structure}(f) is a $\{113\}$. If laterally extended, it could be matched by a $\{335\}$ facet growing out of the $\left(111\right)$ substrate to form a high symmetry boundary. The $\{113\}/\{335\}$ boundary was estimated to have an energy about four to 20 times larger than that of a $\{111\}/\{111\}$ boundary \cite{fullman50,fullman51a,fullman51b}. In other words the $\{111\}/\{115\}$ boundary dissociation also implies the formation of a costly $\{113\}/\{335\}$ replacing a $\{111\}/\{111\}$ boundary parallel to the substrate surface. Yet, the energetic cost is limited due to the short length of the $\{113\}/\{335\}$ boundary and appears compensated by the formation of the other $\{111\}/\{111\}$ boundary (the side one).


Although the dissociated $\{111\}/\{115\}$ structure is energetically more favorable than the undissociated one, it is considerably less favorable than the coherent $(111)$ twin boundary. Therefore upon annealing the driving force to remove the dissociated $\{111\}/\{115\}$ boundaries is much larger than the one for the removal of the coherent twin planes. Thus first the length and extension of the $\{111\}/\{115\}$ boundaries diminishes through coarsening as observed by STM and SXRD, and only then, at higher temperature the twins are eventually removed. We speculate that the removal of the twins taking place at about half of the melting temperature in itself is linked to the onset of bulk diffusion through the onset of vacancy generation.

Considering one twin crystallite, the discussion was focussed on two of its boundaries with the remaining crystal: the twin boundary in the $(111)$ plane, and one of the side boundaries. However, minimization of the system's energy may require maximum contact of the twin grain to its three dimensional matrix, noticeably through contact planes such as the one for the dissociated $\{111\}/\{115\}$ boundary. This may for instance imply strain fields in the grain and surrounding. Such an issue is beyond the scope of the article.


\section{Conclusion \& Summary}

Homoepitaxial thin Ir films growth at room temperature on Ir(111) is characterized by a proliferation of twins starting from 10 ML, leading to an areal fraction of twin approaching 70 \% for about 70 ML, as determined by SXRD and STM. Quantitative analysis of SXRD data from relatively thick films (several 10 ML) was accordingly developped and the twin/fcc composition of the film was obtained. The boundaries between different stacking areas are identified to consist of $\{111\}/\{115\}$ boundaries probably dissociated into coherent $\{111\}/\{111\}$ and $\{552\}/\{112\}$ boundaries. This transformation takes place on the $\{1 1 1\}$ side-facets. It involves the formation of a volume twinned with respect to twin crystallites. The proposed structure of the dissociated boundaries is consistent with the observed additional peaks in the the SXRD data and the STM observation of fractional step heights of 1/3 and 2/3 of a monolayer along $\left\langle 1 \bar{1}0\right\rangle$.

During annealing of the Ir thin films first the area of the boundaries separating different stacking diminishes through coarsening at temperatures in the range of 800-1000\,K. Only at much higher temperatures beyond 1200\,K also the twins themselves heal. The two step annealing process is traced back to a difference in the driving force for healing: the boundaries between different stacking domains are energetically more costly than the coherent twin boundaries.

\begin{acknowledgments}
Support by the Deutsche Forschungs\-gemein\-schaft through the project ``Kinetics of stacking faults in thin films'' is acknowledged.
\end{acknowledgments}

\appendix

\section{Details of the simulation of the X-ray intensity scattered along CTRs}

The X-ray attenuation length was taken into account in the calculation through an imaginary part \cite{henke93} for the scattering vector resulting form the absorption index which adds to the refraction index of X-rays.

\begin{figure}
\begin{center}
\includegraphics[width=0.8 \linewidth]{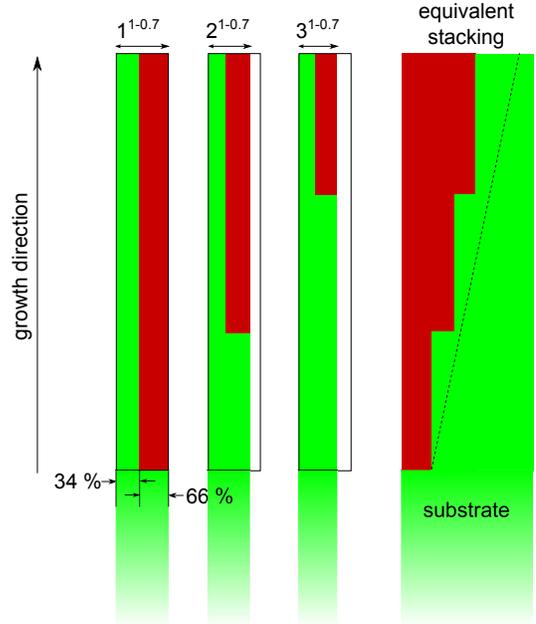}
\end{center}
\caption
{(Color online) Schematics of the model used for the X-ray scattering simulations. The twinned film is sliced in 10 columns (for the sake of clarity this is exemplified here with three). Each of the columns comprises fcc and twinned stacking, in light gray (green) and dark gray (red) respectively. As the index of the column increases, the first twinned regions (coexisting with fcc ones in a $f/{1-f}$ proportion with $f={66 \over 100}$) are found higher in the column. Each column is given a different lateral size, which is accounted for by the exponent $\eta=0.7$ (see text). The equivalent stacking along the growth direction is built by arranging twin regions from each column side by side, and the same for fcc regions. The distinct lateral sizes of each column result result in a non-linear (dashed line) evolution of the twin fraction as a function of the height in the thin film.}
\label{modelcalc}
\end{figure}

The calculation includes the contributions from the bulk and the thin film on top, as well as the interferences between them. Since the lateral extension of the twin and fcc regions, as assessed by STM, is much smaller than the X-ray beam coherent length, their interferences are included as well. For the calculation, 10 vertical domains (columns) are considered. The twin nucleation at different stages of the thin film growth is accounted for as follows. In each of the 10 domains one stacking fault, initiating the formation of a twin, is formed at a given height. If $i$ is the index of the domain, $i\in \left[1,10\right]$, and $t$ is the film thickness, the stacking fault is formed at a height $i\times ip\left(t\over 10\right)$, $ip$ being the integer part function. Below this height the domain consists in a pure fcc crystal, above it consists in a fraction $f$ of twin and the remaining of fcc (see Fig. \ref{modelcalc} for an illustration with three domains). If all domains were given the same weight in the summation of their structure factors, the overall fcc/twin fraction would increase linearly along the growth direction (inclined dashed line in Fig. \ref{modelcalc}). So that this increase can be slower or faster, the domains may all have a different weight, \textit{i.e.} a different lateral size. Instead of using independent weights, which would imply 10 parameters to be refined along the fit procedure, we impose that the weight $W_i$ of a domain depends on its index, \textit{i.e.} $W_i = i^{1-\eta}$, $\eta$ being the only parameter to be refined. This is exemplified in Fig. \ref{modelcalc}.

From this model one may deduce the equivalent twin/fcc fraction at a given height in the thin film by adding the fcc and twin contributions from each of the 10 domains at this height. For the atomic layer labelled $n$ in the film ($n=1$ for the layer in contact with the substrate), the fcc/twin fraction is constant if $n\in\left[ip\left(t\over 10\right)\times \left(j-1\right), ip\left(t\over 10\right)\times j\right]$ with $j\in\left[1,10\right]$ since new twinned domains nucleate only at multiples of $ip\left(t\over 10\right)$. So for $n\in\left[ip\left(t\over 10\right)\times \left(j-1\right), ip\left(t \over 10\right)\times j\right]$, the twin fraction is
\begin{equation}
T_n =\frac{f \sum_{k=1}^j k^{1-\eta} }{S}
\end{equation}
and the fcc fraction is
\begin{equation}
F_n = \frac{\left(1-f\right) \sum_{k=1}^j k^{1-\eta} + \sum_{k=j+1}^{10} k^{1-\eta}}{S}
\end{equation}
with the normalization factor
\begin{equation}\begin{array}{ccl}
S & = & \left(1-f\right) \sum_{k=1}^j k^{1-\eta} + \sum_{k=j+1}^{10} k^{1-\eta } +f \sum_{k=1}^j k^{1-\eta}\\
& = & \sum_{k=1}^{10} k^{1-\eta}
\end{array}\end{equation}

Figure \ref{ctrs}(h) was derived from these equations.

Note that the choice of 10 vertical domains (and thus of stacking fault nucleation at multiples of $ip(t/10)$ in each domain) is a compromise between computation time and fine description of the system. The use of more than 10 domains would have been prohibitively time consumming along with the fitting procedure, which implies calculating the intensity a large number of times.

The roughness is included following a $\beta$-model, namely the film does not end abruptly but rather a number of layers are added on top which occupancy decays with the index number $n$ of the layer following $\beta^n$, $\beta \in \left] 0,1 \right[$ being the roughness parameter. The $\beta$ parameter is chosen the same for the twin and fcc regions. $\beta$ may be converting into an rms roughness by calculating ${\beta\over{1-\beta}}d_{Ir}$, $d_{Ir}$ being the lattice spacing perpendicular to the surface. It should be noted that we found that translations of some regions in the crystal ensure the crystallite contacts at twin boundaries. These translations have a component perpendicular to the substrate corresponding to shifts of a fraction of the lattice spacing. In addition, the decoration rows \cite{bleikamp08} also are regions of the crystal into which the atomic positions are displaced perpendicular to the surface with respect to the atomic positions in fcc and twin regions. Such contributions (fractionnal shifts and decoration rows) of a non integer multiple of $d_{Ir}$ are included in the rms roughness evaluation by STM, and not in the model for the simulation of the CTRs scattered intensity. This makes the direct comparison between the values derived by the two techniques very unreliable.

Four parameters are refined during a fit: the twin fraction $f$, the exponent $\eta$ for the domains weights, the $\beta$ roughness parameter, plus a scale factor between the data and the simulation. The fit is performed after the data is corrected taking into account the Z-axis geometry \cite{Robach2000} of the goniometer which is relevant here.

\section{Details of the simulation of the contribution of $\{111\}/\{115\}$ twin boundaries to the X-ray scattered intensity}

For simplicity, the calculation was here performed without taking into account any attenuation length for the X-rays. The structure factor was computed for a three dimensional crystal including a fcc part and the remaining twinned along $\left[111\right]$. The side of the fcc part closeby the twin one ends with a $\{115\}$ facet and the corresponding facet for the twin is a $\{111\}$. Two situation were considered, with or without a 1/3 $\left[255\right]$ translation thanks to which the fcc and twin parts may be brought in contact. For the two-dimensional maps of the scattered intensity presented in this article, the dimensions of the overall crystal (twin+fcc) were ~130\,\AA~along $\left[111\right]$, and 110\,\AA~along both $\left[\bar{2}11\right]$ and $[\bar{1}\bar{1}2]$. The generality of the qualitative conclusions drawn in the text do not depend on the relative fraction of twin and fcc or size of the crystal.

Since the crystal's ends are perfectly abrupt, finite size fringes develop in the intensity maps.

\bibliography{paperIV_ir111_ausheilen}

\begin{thebibliography}{32}
\bibitem[{foo()}]{footnote1}
\bibinfo{note}{Note that there are four possible stacking fcc stacking
  sequences, three with a fault underneath.}
\bibitem[{not()}]{notonroughness}
\bibinfo{note}{This is also observed in the SXRD data, namely the integrated
  intensity measured in anti-Bragg position, displays ML oscillations (not
  shown here), consistent with a layer-by-layer growth, up to 10 ML}.
\end{thebibliography}

\end{document}